# Incommensurability and atomic structure of c(2x2)N/Cu(100)

T. Choi, C. D. Ruggiero, and J. A. Gupta

Department of Physics, The Ohio State University, Columbus, OH 43210, USA

## Abstract

We use a scanning tunneling microscope operating in a low temperature, ultrahigh vacuum environment to study the atomic structure of single layer films of $Cu_2N$ grown on Cu(100). The *c*(2x2) lattice of $Cu_2N$ is incommensurate, with a lattice constant of $3.72 \pm 0.02$ Å that is 3% larger than the bare Cu(100) surface. This finding suggests that strain due to lattice mismatch contributes to self assembly in this system. We find that the image contrast on $Cu_2N$ islands depends on bias voltage, which reconciles several interpretations in the literature. We assign features in these STM images to the Cu, N and hollow sites in the $Cu_2N$ lattice with the aid of co-adsorbed CO molecules. This atomic registry allows us to characterize four different defects on $Cu_2N$, which influence the sticking coefficient and electronic coupling of adsorbates.

Ultrathin insulating films enable nanoscale control over the coupling of adsorbates to the surface electron density of metal substrates [1,2,3]. For example, adsorbates on insulating films experience reduced hybridization, allowing direct imaging of molecular orbitals similar to the free molecule [1], and spin-flip spectroscopy of single atoms [2]. Tunneling spectroscopy indicates that despite being only one monolayer thick, films of adsorbed nitrogen on Cu act as an insulator with a band gap exceeding 4 eV [4]. This property has been exploited to decouple the magnetic moment of atomic-scale structures from surface electron density [3].

Adsorbed nitrogen on Cu(100) self assembles on two distinct length scales. At low coverage, nitrogen atoms self assemble into small, irregular islands which exhibit a $c$(2x2) lattice. We refer to these islands as 'copper nitride' or $Cu_2N$. With increasing coverage and annealing, the islands become nearly square-shaped, with an average area of ~25 ± 5 nm$^2$ [5,6]. The islands themselves self assemble into a grid-like array which has attracted interest for nanoscale templating [7]. At saturation coverage, the islands coalesce into quasi-continuous, monolayer films which exhibit trench defects. Despite considerable interest in this system, the mechanisms for self assembly are not well understood. Electrostatic interactions due to charge transfer [8,9], work function differences [10], strain due to an incommensurate lattice [5], intrinsic differences in surface stress tensors [11], and other surface relaxations [12] have been considered as factors contributing to the self assembly in this system.

While scanning tunneling microscopy (STM) is useful for identifying assembly mechanisms, the interpretation of STM images of $Cu_2N$ has also been debated. Based on co-adsorption studies, Leibsle et al., have proposed that the STM images N atoms as protrusions [5]. In contrast, Driver et al., have proposed a rumpling model based on STM images showing two distinct sets of protrusions with (1x1) periodicity assigned to Cu atoms [13]. A difficulty with both these room-temperature STM studies, is the uncertain termination of the tip. More recently, in low-temperature STM measurements by Hirjibehedin et al., protrusions were assigned to hollow sites between Cu and N atoms, based on a lattice required to consistently account for island boundaries [3].

Here we present STM measurements of the $Cu_2N$/Cu(100) system which address the longstanding uncertainties in atomic structure and imaging contrast. After careful calibration, our STM images indicate that the lattice of $Cu_2N$ is not commensurate with Cu(100). This suggests that strain due to lattice mismatch contributes to the self-assembly of islands in this system. Our STM images show a reversal in contrast with voltage, which reconciles the disparate interpretations in Refs 3,5,13. Using co-adsorbed carbon monoxide molecules, we can assign features in our images to the Cu and N lattices. This atomic registry is used to characterize four different types of defects commonly seen in the $Cu_2N$ islands.

All measurements were made with a Createc UHV LT-STM which operates at a temperature of 5.5 K in an ultrahigh vacuum environment ($< 1\times10^{-10}$ mbar). Under these conditions, thermal drift and sample contamination are negligible. The bias voltage, $V$, refers to the sample voltage. A cut Ir tip was prepared with field emission and controlled contact with the sample. The Cu(100) surface was prepared by repeated $Ar^+$ sputtering and annealing cycles (~600 °C). Auger electron spectroscopy was used to monitor sample cleanliness and N absorption. $Cu_2N$ islands were grown by sputtering the clean Cu sample in a $N_2$ atmosphere ($1\times10^{-5}$ mbar) for 2 min. The high voltage of the sputter gun dissociates $N_2$, so that N atoms are deposited onto the Cu surface. After nitrogen adsorption, the sample is annealed to 350ºC for 1 min. $Cu_2N$ islands were isolated on the surface, at an average coverage of 0.16 ML. The sample is then cooled to 80 K and inserted into the cold STM. Carbon monoxide molecules are co-adsorbed onto the surface in the STM at 12 K through a leak valve. CO molecules are identified by before/after images of the same area, and inelastic electron tunneling spectroscopy of the characteristic frustrated rotation and translation modes [14].

We calibrated our STM images by measuring the lattice constant of three ~50 $nm^2$ regions of bare Cu(100), located between $Cu_2N$ islands. To resolve metal atoms in close-packed surfaces such as Cu(100), the STM tip must be brought nearly into point contact (R ~ 100 kΩ). Thus, under typical imaging conditions (R>10 MΩ), the lattice of Cu atoms is not directly observed. We instead deliberately transfer a coadsorbed CO molecule from the surface to the tip by applying a voltage pulse. Such a functionalized tip exhibits enhanced contrast as a result of chemical interactions between the molecule and the surface [15]. With a CO-terminated tip, we are readily able to image the (1x1) lattice of Cu(100) [16]. The apparent lattice constant was measured from Gaussian fits to the spot profile in a 2D-FFT [17]. Images were then scaled assuming a lattice constant of 2.55Å for Cu(100). We note that photoemission measurements by Sekiba et al., suggest that bare regions of Cu(100) contract by 0.5-1.2%, depending on nitrogen coverage [18]. Such a contraction would proportionately decrease the $Cu_2N$ lattice constant we report, but would not otherwise affect our conclusions. Care was taken to ensure constant temperature after calibration, because the sensitivity of the piezoelectric tube used for scanning exhibits hysteresis of a few percent after temperature cycling.

Figure 1A shows an STM image of a $Cu_2N$ island. A series of bright protrusions corresponding to the $c$(2x2) lattice of $Cu_2N$ can be resolved by a metal tip under typical imaging conditions (e.g. R > 100 MΩ). We measured the lattice constant of 26 $Cu_2N$ islands ranging in area from 6 nm$^2$ to 25 nm$^2$ by 2D-FFT as described above. Our error in these measurements is mainly due to the number of protrusions within the $Cu_2N$ islands. We find no variation of lattice constant with island size in this range. The weighted average from these measurements is $a_{Cu2N}$ = 3.72 ± 0.02 Å, which lies between that of the closest bulk counterpart, $Cu_3N$ (3.8 Å), and bare Cu (3.61 Å).

The ratio $a_{Cu2N}$ / $a_{Cu}$ = 3.72/2.55 = 1.46 ± 0.01 indicates that the $Cu_2N$ lattice is *incommensurate*, independent of any corrections due to contraction of the Cu(100) lattice [18]. If $Cu_2N$ were commensurate, the ratio would be $\sqrt{2}$ = 1.414. Therefore, surface strain due to lattice mismatch should contribute to self assembly in this system, a suggestion first made by Leibsle et al. [5]. This has been a point of contention in the literature, due in part to significant uncertainty about the exact atomic structure of $Cu_2N$, which could not be quantified in the early STM measurements. A LEED study suggested a perfectly commensurate structure [19], although subsequent Rutherford channeling studies by the same group suggested a significant (~ 0.1 Å) vertical and lateral expansion of the Cu lattice in $Cu_2N$ [11,20]. Density functional theory [8, 9] and x-ray measurements suggest that nitrogen atoms sit above the plane of Cu atoms, although measured heights vary from 0-0.6 Å [5]. The uncertainty in nitrogen height translates to an uncertainty in the $Cu_2N$ lattice constant; x-ray measurements report a bond length of $d_{Cu-N}$ = 1.85 Å [21], which corresponds to 3.5 Å< $a_{Cu2N}$ <3.7 Å, depending on the nitrogen height one chooses. We do not directly measure $d_{Cu-N}$ in our STM measurements, but our value of $a_{Cu2N}$ / 2 = 1.86 Å sets a lower limit for this value, attained assuming co-planar N and Cu atoms.

Figures 1A-C show STM images of the same $Cu_2N$ island at different sample voltage. Inspection of these images reveals that the series of protrusions which exhibit the $c$(2x2) periodicity depend on the voltage. This is clearly seen in the insets, which show exactly the same area of the island at each voltage. Depressions at positive voltage become protrusions at negative voltage, and vice versa. At low voltage (Fig. 1B), we observe two sets of protrusions, defining a (1x1) lattice.

Exact registry of Cu atoms in the bare surface, with the features in $Cu_2N$ islands is needed to interpret these images. STM images with simultaneous atomic resolution on Cu and $Cu_2N$ surfaces provide the simplest way to obtain this registry [6,16,22]. However, in our experience, simultaneous atomic resolution is only possible with an adsorbate-terminated tip. While useful for calibration, we believe such tips are not reliable for assigning adsorption sites, because the chemical contrast mechanism can vary with the surface and tip adsorbate.

We instead co-adsorb a well studied reference adsorbate to determine binding sites in the $Cu_2N$ islands [5]. Carbon monoxide (CO) molecules are known to adsorb atop Cu atoms on Cu(100) [14]. Five CO molecules are situated near the $Cu_2N$ island in Figure 1. The Cu(100) lattice (black circles) is inferred from the position of these CO molecules, with a spacing and rotation as determined in our calibration measurements of bare Cu(100) regions.

Lattice mismatch complicates the registry of features in the STM images of $Cu_2N$ with the Cu(100) substrate [23]. We assign the depressions at positive voltage (Fig. 1A) to nitrogen atoms, based on two criteria:

1) *Proximity to hollow sites:* experiment and theory agree that nitrogen atoms adsorb at the four-fold symmetric hollow sites on Cu(100) [5,8,9]. If we extend the Cu lattice over the $Cu_2N$ island in Fig. 1A, both depressions and protrusions in the (2x2) lattice lie nearest the hollow sites [23]. *A priori*, either one of these could be assigned to nitrogen atoms.

2) *Consistency with island boundaries:* We distinguish between these choices by considering the island boundary [3]. At positive voltage (Fig. 1A), atomically sharp depressions form corners along the islands' perimeter (arrows in Fig. 1A). In STM images where we inhibit island formation by reduced annealing temperature, we find that individual nitrogen atoms are imaged as depressions at positive voltage. Thus, we locate the lattice of nitrogen atoms as needed to consistently account for the sharp corners in Fig. 1A.

The lattice of Cu atoms is chosen in a similar fashion. In this case, if we extend the Cu lattice onto the $Cu_2N$ island, imaged at positive or negative voltage, Cu sites lie nearest the bridge positions in the (2x2) lattice [23]. This is consistent with the required (1x1) periodicity for features assigned to the Cu atoms.

We thus define the $Cu_2N$ lattice shown in Figs. 1A-C. Protrusions at negative voltage (Fig. 1C) locate nitrogen atoms, while protrusions at positive voltage (Fig. 1A) locate hollow sites between Cu and N atoms. This lattice assignment, together with our observation of voltage-dependent contrast, reconciles the interpretations in Refs. 3,5,13. The assignment of protrusions at negative voltage to nitrogen atoms agrees well with DFT calculations [8,24], although the reversal in contrast we observe was not reproduced in the calculations. Prior STM [6] and DFT studies [24] suggest a nonuniform expansion of the $Cu_2N$ (2x2) lattice. However, the uniform lattice shown in Figure 1 fits well to the entire island. We suggest that any lattice distortion is limited to regions near the boundaries, which do exhibit distinct contrast in STM images.

Figure 1B shows a more complicated (1x1) periodicity at low voltage. To eliminate possible artifacts which vary with tip height, we adjusted the set current so that the tip height over $Cu_2N$ was identical in Figs. 1B, C. From our lattice assignment, we see that both nitrogen and hollow sites are imaged simultaneously as protrusions at low voltage. Such images were first observed by Driver et al., who developed a rumpling reconstruction model in which both sets of protrusions were assigned to Cu atoms [13]. Our measurements instead indicate that this is an electronic contrast effect.

We note a correspondence of these images with voltage-dependent contrast in GaAs(110) [25]. There, it was observed that the STM images occupied valence states (i.e. As atoms) at negative voltage, and unoccupied conduction states (Ga atoms) at positive voltage. In our STM measurements of insulating $Cu_2N$ [4], valence states are associated with nitrogen atoms, and conduction states are associated with the empty hollow sites, perhaps reflecting leakage of state density from the underlying Cu substrate.

Defects play an important role in the properties of few-monolayer insulating films. Not only can defects define conducting channels through the insulating film, they can also affect the interaction between surface and adsorbates. For example, oxygen vacancies on MgO films promote charge transfer to adsorbed metal clusters, which influences their catalytic activity [26]. Adsorbates may also preferentially stick to these defects, due to the higher electron density available for bonding.

To characterize these effects, we adsorbed an organic molecule, azobenzene ($C_{12}H_{10}N_2$), on the $Cu_2N$/Cu(100) surface. Consistent with studies in other systems [27], we find that azobenzene preferentially sticks to exposed Cu regions, with a ratio of ~ 25: 1. The few molecules we find on $Cu_2N$ islands are typically tethered to defects in the $Cu_2N$ film. For example, Figure 2A shows an STM image of an azobenzene molecule on a $Cu_2N$ island. This molecule was transferred to the STM tip by applying a voltage pulse. Subsequent imaging with the molecule-terminated tip indicates that the azobenzene molecule was adsorbed atop a defect in the Cu$_2$N film (Figs. 2B,C). The spatial resolution in these images is improved by the molecule-terminated tip, although the electronic contrast remains identical to the bare tip images (Fig. 1).

While on average we observe < 0.3 defects per island, the island in Figure 2 happened to have all four of the defects we commonly observe. Using our lattice assignment, we propose origins for two of the defects. For clarity, we refer to the appearance of these defects at negative voltage (Fig. 2C). We attribute a twofold symmetric depression centered on Cu sites to a Cu vacancy (right inset). We observe three different fourfold symmetric defects centered at the nitrogen site. We attribute a ~0.2 Å protrusion (center inset, top) to a N vacancy, based on our observation that such defects can be created by applying a voltage pulse > 3 V to the STM tip. We are not able to resolve a nitrogen atom on the surface after this procedure, but it is likely that the atom desorbs completely, or falls back to the surface away from our scan area. The two other, less common defects at the nitrogen lattice site appear as a ~0.3 Å depression (left inset), and a higher ~0.6 Å protrusion (center inset, bottom). These may be substitutional defects, or associated with subsurface Cu impurities underneath the Cu$_2$N island.

These STM measurements address some of the debate concerning the self-assembly and interpretation of STM images in the Cu$_2$N/Cu(100) system. The *c*(2x2) lattice is shown to be incommensurate, with a lattice constant of 3.72 ± 0.02 Å. The variation in image contrast with voltage reconciles the disparate interpretations in prior STM studies, and allows us to assign a lattice of Cu and N atoms within the islands.

**Acknowledgements:** We are grateful to the NSF for support through CAREER award DMR-0645451.

# Choi et al., Figure 1

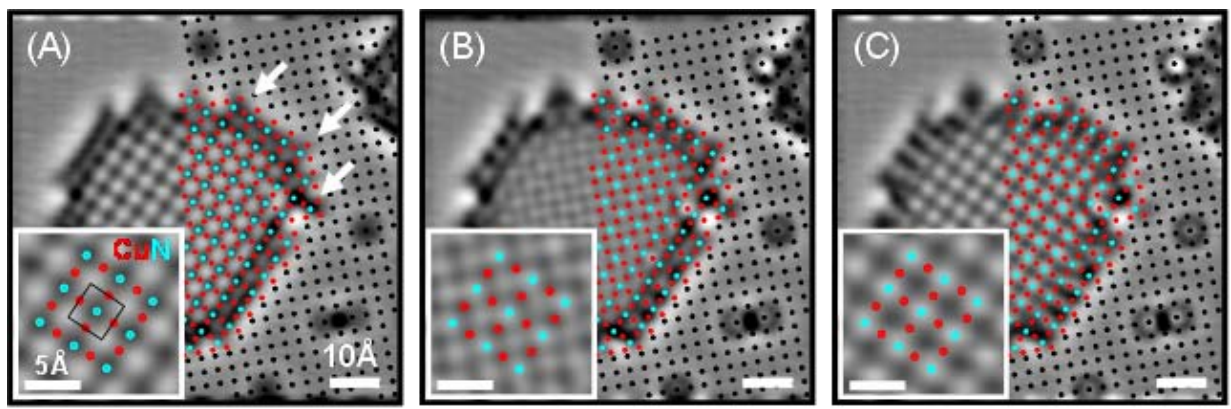

Fig. 1. (color online) STM images of a $Cu_2N$ island on Cu(100). To emphasize local contrast, the images were low-pass and Laplace filtered. Black dots indicate the lattice of copper atoms on Cu(100), inferred from the positions of marker CO molecules. Red dots show copper atoms within the $Cu_2N$ island, and blue dots represent nitrogen atoms, according to the lattice assignment described in the text. Arrows indicate boundary features used to assign the nitrogen lattice. Insets show a higher magnification image of the island center, and the unit cell. (a) +0.5V, 2nA. (b) -0.1V, 1.5nA. (c) -0.5V, 6nA.

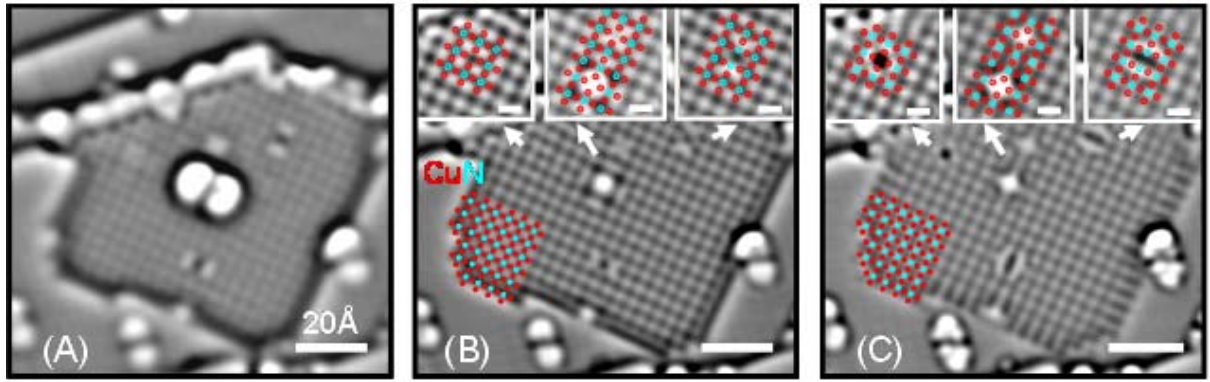

Fig. 2. (color online) STM images of defects in a Cu$_2$N island. (a) An azobenzene molecule adsorbed on a Cu$_2$N island with several defects. V=+0.5 V, I= 0.1 nA. (b-c). The molecule is transferred to the tip with a voltage pulse, revealing an underlying defect. The molecule-terminated tip increases resolution, but does not change the electronic contrast. Insets show four typical defects (scale bar = 5 Å), together with the assigned Cu$_2$N lattice. (b) +0.5 V, 0.1 nA. (c) -0.5 V, 0.1 nA.

# Choi et al., Supplemental Figure 1

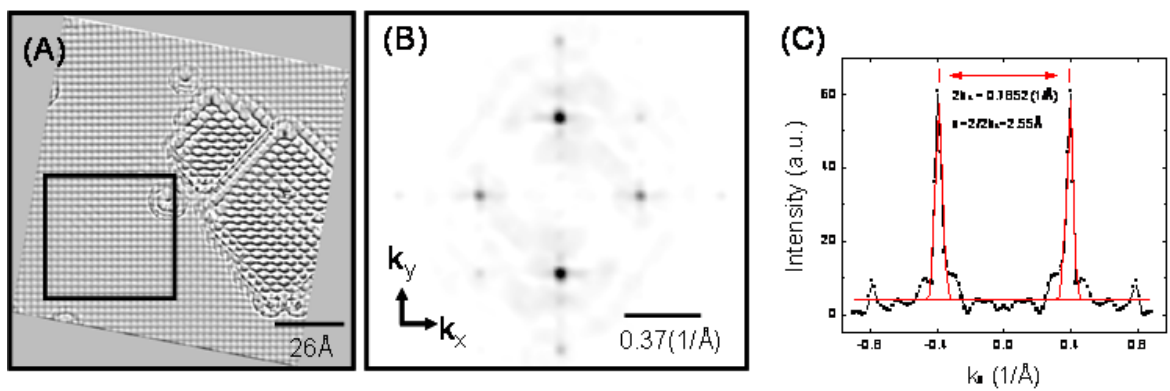

Fig. S1. (A) STM image (+25mV, 2.4nA), with simultaneous atomic resolution on bare Cu(100) and $Cu_2N$ islands using a CO-terminated tip. The image was rotated in software to align the Cu lattice, and calibrated based on analysis of the corresponding 2D FFT in (B). (C) Gaussian fit to the FFT spot profile. The computed value of 2.55 Å for Cu(100) confirms the calibration of the STM image in A.

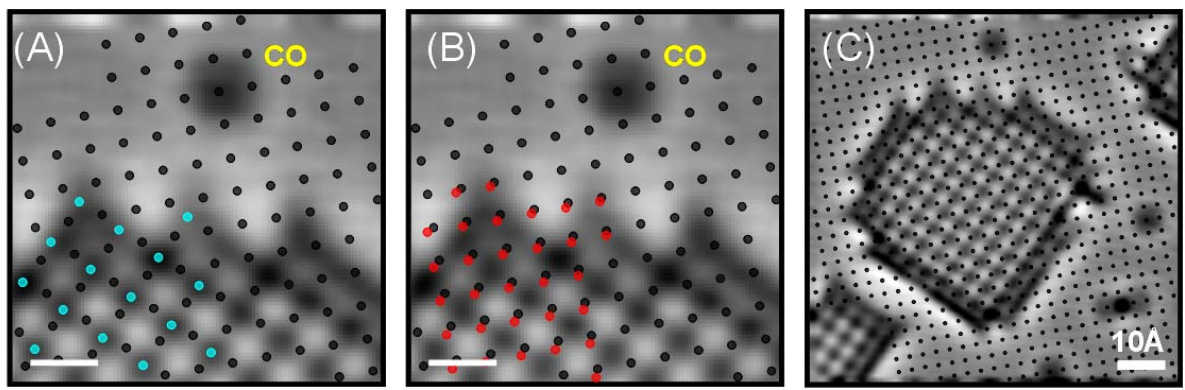

Fig. S2. (A-B) High magnification images (scale bar = 5 Å) for assignment of N and Cu lattices in $Cu_2N$ islands. V=+0.5V, I=2nA. The inferred Cu(100) lattice (black points) is extended onto the $Cu_2N$ island. (A) The N lattice (blue points) is assigned to depressions based on proximity to Cu hollow sites, and consistency with sharp corners in the island. (B) The Cu lattice (red points) is assigned to bridge sites based on proximity to the Cu(100) lattice. (C) Direct indication of incommensurability: registry between the Cu(100) lattice and protrusions or depressions in the c(2x2) lattice changes over the island. Same image as Fig. 1A.

# References

[1] J. Repp, G. Meyer, S.M. Stojković, A. Gourdon, and C. Joachim, Phys. Rev. Lett. **94**, 026803 (2005).

[2] A.J. Heinrich, J.A. Gupta, C.P. Lutz, and D.M. Eigler, Science **306,** 466 (2004).

[3] C.F. Hirjibehedin, C.P. Lutz, and A.J. Heinrich, Science **312**, 1021 (2006).

[4] C.D. Ruggiero, T. Choi, and J.A. Gupta, Appl. Phys. Lett. **91**, 253106 (2007).

[5] F.M. Leibsle, S.S. Dhesi, S.D. Barrett, and A.W. Robinson, Surf. Sci. **317**, 309 (1994).

[6] F. Komori, S. Ohno, and K. Nakatsuji, Prog. Surf. Sci. **77**, 1 (2004).

[7] T.M. Parker, L.K. Wilson, N.G. Condon, and F.M. Leibsle, Phys. Rev. B **56**, 6458 (1997).

[8] Y. Yoshimoto and S. Tsuneyuki, Surf. Sci. **514**, 200 (2002).

[9] A. Soon, L. Wong, B. Delley, and C. Stampfl, Phys. Rev. B **77**, 125423 (2008).

[10] D. Vanderbilt, Surf. Sci. **268**, L300 (1992).

[11] B. Croset, Y. Girard, G. Prevot, M. Sotto, Y. Garreau, R. Pinchaux, and M. Sauvage-Simkin, Phys. Rev. Lett. **88**, 056103 (2002).

[12]. H. Ellmer, V. Reparin, S. Rousset, B. Croset, M. Sotto and P. Zeppenfeld, Surf. Sci. **476**, 95 (2001).

[13] S.M. Driver and D.P. Woodruff, Surf. Sci. **492**, 11 (2001).

[14] L.J. Lauhon and W.Ho, Phys. Rev. B **60**, R8525 (1999).

[15] H.J. Lee and W. Ho, Science **286**, 1719 (1999).

[16] See EPAPS Document No. [x] for images showing the analysis procedure.

[17] I. Horcas, R. Fernandez, J.M. Gomez-Rodriguez, J. Colchero, J. Gomez-Herrero, and A.M. Baro, Rev. Sci. Instr. **78**, 013705 (2007).

[18] D. Sekiba, Y. Yoshimoto, K. Nakatsuji, Y. Takagi, T. Iimori, S. Doi, and F. Komori, Phys. Rev. B **75**, 115404 (2007).

[19] M. Sotto and B. Croset, Surf. Sci. **461**, 78 (2000).

[20] C. Cohen, H. Ellmer, J.M. Guigner, A. L'Hoir, G. Prevot, D. Schmaus, and M. Sotto, Surf. Sci. **490,** 336 (2001).

[21] T. Lederer, D. Arvanitis, M. Tischer, G. Comelli, L. Tröger, and K. Baberschke, Phys. Rev. B **48**, 11277 (1993).

[22] T.E. Wofford, S.M. York, and F.M. Leibsle, Surf. Sci. **522**, 47 (2003).

[23].See EPAPS Document No. [x] for images showing the lattice registration procedure.

[24] Y. Yoshimoto and S. Tsuneyuki, Appl. Surf. Sci. **237**, 274 (2004).

[25] R.M. Feenstra, J.A. Stroscio, J. Tersoff, and A.P. Fein, Phys. Rev. Lett. **58**, 1192 (1987).

[26] B. Yoon, H. Hakkinen, U. Landman, A.S. Worz, J.M. Antonietti, S. Abbet, K. Judai, and U. Heiz, Science **307**, 403 (2005).

[27] N. Nilius, T.M. Wallis, and W. Ho, Phys. Rev. Lett. **90**, 046808 (2003).